\documentclass[twocolumn,prl]{revtex4}
\usepackage{hyperref}
\usepackage{bm,amsmath}
\usepackage{graphicx,psfrag}
\usepackage{color}
\usepackage{amsmath, latexsym, amsfonts, amssymb, amsthm, amscd}
\newcommand \be  {\begin{equation}}  
\newcommand \bea {\begin{eqnarray} \nonumber }  
\newcommand \ee  {\end{equation}}  
\newcommand \eea {\end{eqnarray}}  

\begin{document}

\title{Critical dynamical heterogeneities close to continuous second-order glass transitions}
\author{Saroj Kumar Nandi$^{1}$, Giulio Biroli$^{1}$, Jean-Philippe
Bouchaud$^{2}$, Kunimasa Miyazaki$^{3}$, David R. Reichman$^{4}$}
\affiliation{
$^1$ Institut de Physique Th\'eorique, CEA/DSM/IPhT-CNRS/URA 2306 CEA-Saclay,
F-91191 Gif-sur-Yvette, France}
\affiliation{
$^2$ Capital Fund Management, 23 rue de l'Universit\'e, 75007 Paris, France.}
\affiliation{$^3$ Institute of
Physics, University of Tsukuba, Tsukuba 305-8571, Japan}
\affiliation{$^4$ Department of Chemistry, Columbia University, 3000 Broadway, New York, NY 10027, USA.}

\date{\today}

\begin{abstract}
We analyse, using Inhomogenous Mode-Coupling Theory, the critical scaling behaviour of the dynamical susceptibility at a distance $\epsilon$ from continuous second-order glass transitions. We find that the dynamical correlation length $\xi$ behaves generically as $\epsilon^{-1/3}$ and that the upper critical dimension is equal to six. More surprisingly, we find {\it activated} dynamic scaling, where $\xi$ grows with time as $\ln^2 t$ exactly at criticality. All these results suggest a deep analogy between the glassy behaviour of attractive colloids or randomly pinned supercooled liquids and that of the Random Field Ising Model.
\end{abstract}


\maketitle

Several recent studies have revealed that the properties of the glass transition can be 
drastically modified by suitably tuning some control parameters. In the case of the colloidal glass 
transition, an attractive interaction on top of the hard-sphere repulsion can change the dynamical behaviour and lead to glass-glass transitions and logarithmic relaxation \cite{sciortino}. This behaviour is also expected for glassy liquids in porous media \cite{krakoviack}. 
For generic glass-forming liquids, it was recently predicted \cite{CBRPGT,CC} that randomly pinning a fraction of particles would transmute the glass transition in a continuous second-order phase transition akin to that of the Random Field Ising model (RFIM) \cite{CBRPGT, CC, FPRFIM}. There are strong theoretical indications that these phenomena are in fact all related to the existence of a new kind of glassy critical point, 
first found within Mode-Coupling Theory (MCT) as a higher-order singularity \cite{zaccarelli}. The physical contexts in which it appears are quite different: for attractive colloids it
is a terminal point of a glass-glass transition line, for glass-forming liquids either pinned or trapped in porous media it corresponds to the locus where 
the Mode-Coupling transition and the Ideal Glass transition lines merge. In the former case the glass transition line stops at this new critical point \cite{CBRPGT}, whereas in the latter it carries on and becomes continuous \cite{krakoviack}. Remarkably, in all these physical situations, activated processes not described by MCT are either suppressed, since the region where they appear shrinks to zero for glass-forming liquids, or not relevant since the system is in the glass phase for attractive colloids. Hence, MCT might become {\it quantitatively} accurate in these situations.  
The dynamical behaviour of the two-point functions at this new glassy critical point, that we will call A3 using G\"otze's terminology, was predicted by MCT computations \cite{zaccarelli} and confirmed later both numerically and experimentally in colloids \cite{sciortino}.
The static properties of the fluctuations of the overlap field between two equilibrium configurations were recently investigated in \cite{CBRPGT, FPRFIM} and, using simulations, in \cite{berthierkob}.  A complete theory of dynamical correlations is however still lacking. The aim of this paper is to develop such a theory by 
extending the ``Inhomogeneous'' MCT (IMCT) formalism, which was developed by some of us \cite{IMCT} to describe 
dynamical heterogeneities at the usual MCT transition. We shall obtain the mean-field values of the critical exponents, the upper critical dimension and derive 
the critical behavior, which turns out to be very different from the usual one. We find in particular {\it activated dynamic scaling}, which strongly bolsters the relationship with the RFIM \cite{CBRPGT, FPRFIM}.  

In order to grasp the main properties of the A3 critical point, it is useful to
focus on the mean-field Landau-like potential $V(f;\mathbf{\varepsilon})$, called the Franz-Parisi (FP) potential in the present
context \cite{FP}. The arguments $\mathbf{\varepsilon}$ and $f$ are, respectively, the vector of all control parameters that can be tuned (e.g. the temperature and the fraction of pinned particles)
and the glassy (non-ergodic) order parameter, which measures how far the dynamics can displace the system away from its initial configuration. For usual glass transitions, the FP potential has a unique minimum $f_0=0$ at high temperatures; it corresponds to a complete loss of memory of the initial condition
as normal in a liquid. A secondary minimum appears for $f=f^* > 0$ below a certain transition temperature $T_c$, see Fig. 1. An important achievement 
of the last decades was to establish that $T_c$ actually coincides with the MCT transition, where locally stable, long-lived amorphous structures, corresponding to the secondary minimum of $V(f;\mathbf{\varepsilon})$, appear.
\begin{figure}
\psfrag{e}{$f$}
\psfrag{d}{$V(f,\epsilon)$}
\psfrag{a}{ \small $A_2$ critical point}
\psfrag{c}{\small $A_3$ critical point}
\psfrag{b}{\small $T<T_c$}
\psfrag{g}{$T>T_c$}
\psfrag{f}{}
\psfrag{l}{$T$}
\psfrag{m}{$c$}
\psfrag{n}{$\epsilon_\perp$}
\psfrag{o}{$\epsilon_{||}$}
\includegraphics[width=8.6cm]{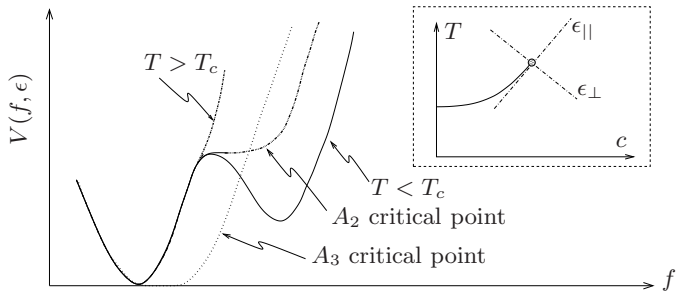}
\caption{Schematic presentation of the FP potential. The second minimum that develops below the transition point becomes flat as the usual MCT ($A_2$) transition point is approached from below. The two minima coalesce one at the $A_3$ point. Inset: schematic phase diagram obtained for random pinning glass transitions. In the temperature-pinning fraction plane the MCT transition line ends in an A3 critical point \cite{CBRPGT}. 
The two special directions mentioned in the text are shown. }
\label{mct_A3_pot}
\end{figure}
Roughly speaking, the connection between the FP potential and MCT may be expressed as 
$V'(f;\mathbf{\varepsilon})=\frac{f}{1-f}-\mathcal{F}_{\mathbf{\varepsilon}}[f]$, where $\mathcal{F}_{\mathbf{\varepsilon}}[f]$ is the memory
function of the MCT equations (The full-fledged MCT calculation deals with wave-vector dependent order parameters $f_{\vec q}$, we will return to this below) \cite{F1}.
An A3 critical point corresponds to the merging of the minima $f_0$ and $f^*$ \cite{Gotze,FZ}. 
It is of co-dimension 2, much as the liquid-gas critical point, {\it i.e.} one needs to tune at least two control parameters to reach it, as found for systems with quenched pinning sites (see inset of Fig. 1) and for hard sphere systems, where one tunes the short-range attractive interactions and the density.

Technically, the existence of an underlying thermodynamical formulation has been extremely useful to understand that the MCT transition 
is necessarily accompanied by the divergence of a length scale, which governs the spatial extent over which dynamical fluctuations 
are correlated, a feature that was hard to anticipate within the original framework of G\"otze et al. This diverging length scale
is in fact a direct consequence of the vanishing of the {\it curvature} $V''(f^*;\mathbf{\varepsilon})$ of the FP potential at $T_c$ (see Fig. 1) \cite{FPchi,BBchi,MF}.  
The IMCT formalism allowed us to make a series of precise predictions about the space-time scaling of dynamical heterogeneities in supercooled liquids close to $T_c$ \cite{IMCT}. One finds in particular that the dynamical correlation length diverges as $|T-T_c|^{-1/4}$ as the critical point is approached. Although the IMCT predictions are only expected to be correct far enough from $T_c$ below 8 dimensions, many general predictions appear to 
be confirmed, sometimes quantitatively, by large scale computer simulations, see \cite{kim} and refs therein. These developments, and others, strongly support 
a quantitative theory of supercooled liquids built using the mean-field scenario as a starting point, much as 
Curie-Weiss theory provides a foundation for the modern theory of critical phenomena \cite{andreanov,urbani}.

In the following we first explain our results in an informal and simple way based on the behaviour of the FP
potential; we then sketch the complete IMCT derivation. We will denote as $V^{(n)}$ the 
$n$-th derivative of $V(f;\mathbf{\varepsilon})$ with respect to $f$ ($V'=V^{(1)}$, etc.), and $\vec \nabla_\varepsilon V$ the gradient of $V$ with respect to the
parameters. The expansion of $V'(f;\mathbf{\varepsilon})$ around the transition point $f=f_c$, $\vec \varepsilon=0$ reads, with $\delta f=f-f_c$:
\bea
V'(f;\mathbf{\varepsilon}) &\approx& V^{(2)}_c \delta f + \frac12 V^{(3)}_c \delta f^2 + \frac13 V^{(4)}_c \delta f^3 
\nonumber 
\\&+& \vec \nabla_\varepsilon V'_c \cdot \vec \varepsilon
+ \vec \nabla_\varepsilon V^{(2)}_c \cdot \vec \varepsilon \, \delta f + \dots,
\nonumber 
\eea
where we have used that by definition, at the transition, $V'(f_c;0)\equiv 0$. The standard MCT transition (the A2 critical point) occurs when the secondary 
minimum of $V(f)$ just appears, implying $V^{(2)}_c=0$ (see Fig. 1). The next order singularity (A3) occurs
when $V^{(3)}_c$ concomitantly vanishes as well (see Fig. 1). Looking for the new location $f^*$ of the minimum away from the transition, one finds, to leading
order in $\epsilon = |\vec \varepsilon|$: $f^* - f_c \sim \sqrt{\epsilon}$, as is familiar for the A2 case and 
$f^* - f_c \sim \sqrt[3]{\epsilon}$ in the generic A3 case. There is however a subtlety here: since the A3 point requires at least two parameters to
be varied simultaneously, one needs to include the case where the chosen trajectory in parameter space is precisely perpendicular to $\vec \nabla_\varepsilon V'_c$, 
in which case one finds again the weaker singularity $f^* - f_c \sim \sqrt{\epsilon_\parallel}$, where $\epsilon_\parallel$ is the distance to the critical point along that special 
direction. The motivation for the notation $\epsilon_\parallel$ stems from the liquid-glass phase diagram, 
sketched in the inset of Fig. 1 in the case of random pinning glass transitions  \cite{CBRPGT}.  
In general A3 is the terminal point 
of the line of A2 critical points. The special direction found above is the one tangent to the A2 line. In consequence, we also introduce the notation $\epsilon_\perp$ 
for the magnitude of the component of $\vec \varepsilon$ perpendicular to the A2 line (i.e. parallel to $\vec \nabla_\varepsilon V'_c$).  \\
The main idea of IMCT \cite{IMCT} is to perturb the system with a small spatially periodic external potential $\propto \cos(\vec q_0 \cdot \vec x)$, whose spatial profile varies over the length-scale $1/q_0$. The characteristic value of $q_0^*$ at which the external perturbation starts to act differently from a uniform,
$q_0=0$, perturbation allows one to obtain the correlation length of dynamical heterogeneities as $\xi=1/q_0^*$. 
Since all the physics of the slowing down is governed by the vanishing of 
the curvature of the FP potential, the crucial point is to work out how the periodic perturbation (of zero mean) changes this curvature. 
Because the system is rotationally invariant, it is reasonable to assume that the extra contribution to the curvature is $\sim q_0^2$. Therefore, one has:
$$
V''(f^*,\mathbf{\varepsilon}) \approx V^{(3)}_c (f^*-f_c) + \frac12 V^{(4)}_c (f^*-f_c)^2 + \vec \nabla_\varepsilon V^{(2)}_c \cdot \vec \varepsilon + \Gamma q_0^2.
$$
Close to an A2 critical point, $V^{(3)}_c \neq 0$ and $f^*-f_c \sim \sqrt{\epsilon}$, which shows that the characteristic value of $q_0$ beyond which the 
relaxation time substantially changes is $\sim \epsilon^{1/4}$, leading to $\xi \sim \epsilon^{-1/4}$, in agreement with the result of \cite{IMCT}. 
Upon approaching an A3 critical point, $V^{(3)}_c = 0$, leading to $\xi \sim \epsilon_\perp^{-1/3}$ in the generic case, and to $\xi \sim 
\epsilon_\parallel^{-1/2}$ in the special case where $\epsilon_\perp=0$. These results are fully confirmed, and made more precise, by the IMCT analysis that 
we now briefly present.
\begin{figure}
\includegraphics[width=7.8cm]{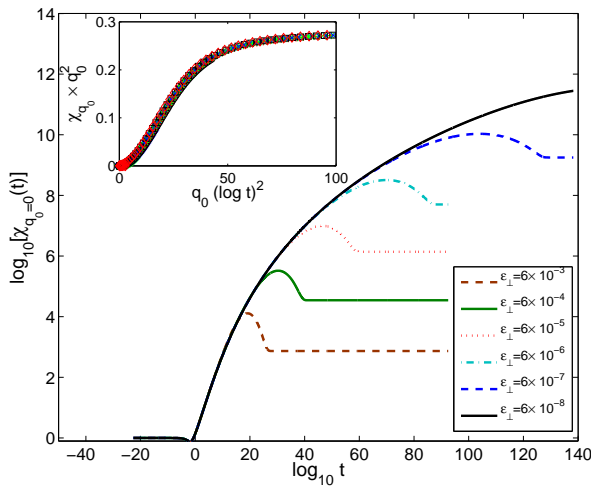}
\caption{Numerical results obtained by solving the schematic $F_{13}$ equations. 
$\chi_{\mathbf{q}_0=0}$ as a function of time approaching the A3 critical point, which is indeed found to rescale as $|\epsilon_\perp|^{-2/3}$ times a 
function of $|\epsilon_\perp|^{1/6}\times \ln t$. Inset: Scaling collapse, for different $q_0$ and $t$, of $q_0^2 \chi^c_{\mathbf{q}_0}$ as a function of $q_0 \ln^2 t$ 
at criticality, as predicted from the theory.}
\label{q0_xi_scaling}
\end{figure}\\
The IMCT formalism starts from an exact equation for the inhomogeneous dynamic structure factor 
$F(\mathbf{q}_1,\mathbf{q}_2,t)= \langle \rho_{\mathbf{q}_1}(t) \rho_{-\mathbf{q}_2}(0) \rangle$ in the presence of an inhomogeneous external field 
$u(\mathbf{q_0})$:
\bea \label{IMCT-eq}
&\frac{\partial F({\bf q}_1,\mathbf{q}_2,t)}{\partial t}+\Omega_{q_1} F({\bf q}_1,\mathbf{q}_2,t)+
\sum_{\mathbf{k}}\int_0^t M(\mathbf{q}_1,\mathbf{k},t-t') 
\\
& \times \frac{\partial F(\mathbf{k},\mathbf{q}_2,t')}{\partial t'} dt'
=\mathcal{T}_u(\mathbf{q}_1,\mathbf{q}_2,t), 
\eea
where $\Omega_{q_1}\equiv q_1^2 k_BT/S_{q_1}$ is a frequency term, $M(\mathbf{q}_1,\mathbf{k},t)$ is the memory kernel and 
$\mathcal{T}_u(\mathbf{q}_1,\mathbf{q}_2,t)$ contains all the terms generated due to the external potential. The dynamical susceptibility 
is defined as $\chi_{\mathbf{q}_0}(\mathbf{q}_1,t)=\delta F({\bf q}_1,{\bf q}_1+\mathbf{q}_0,t)/\delta u(\mathbf{q}_0)|_{u\to 0}$.
This object obeys a linear equation (that we do not write here, see \cite{IMCT}) obtained by taking the derivative of Eq. (\ref{IMCT-eq}) with respect to $u(\mathbf{q}_0)$. 
In the long-time limit, this linear equation reads:
\be 
\label{chi-eq}
\sum_\mathbf{k} \bigg(\delta_{\mathbf{k},\mathbf{q}}- C_{\mathbf{q};\mathbf{k}}(\mathbf{q}_0) \bigg)
\chi_{\mathbf{q}_0}(\mathbf{k},t \to \infty) = S(\mathbf{q},\mathbf{q}_0),
\ee
where $S(\mathbf{q},\mathbf{q}_0)$ is a non singular source term and $C_{\mathbf{q};\mathbf{k}}(\mathbf{q}_0)$ is a $\mathbf{q}_0$ dependent matrix that 
can be fully computed in terms of the memory kernel of the model, see \cite{IMCT}. Note that 
$\chi_{\mathbf{q}_0}(\mathbf{k},t \to \infty)$ is nothing else than the variation of the non-ergodic parameter
$f_\mathbf{k}= F(\mathbf{k},t \to \infty)$ due to the external potential. In order to analyse Eq. (\ref{chi-eq}) we recall (see \cite{Gotze}) that 
$C_{\mathbf{q};\mathbf{k}}(0)=(1-f_\mathbf{k})^2 \partial \mathcal{F}_{\mathbf{q},\mathbf{\varepsilon}}[\{f_\mathbf{k}\}]/\partial f_\mathbf{k}$. The properties of the operator $\delta_{\mathbf{k},\mathbf{q}}- C_{\mathbf{q};\mathbf{k}}(0)$, which is akin to $V'(f;\mathbf{\varepsilon})$, are reported in \cite{Gotze}: at  distance $\epsilon$ from the transition, one finds $f_\mathbf{k}=f^c_\mathbf{k}+(1-f^c_\mathbf{k})^2 g_\mathbf{k}$, where $g_\mathbf{k} \propto \sqrt[3]{\epsilon} \psi^R_\mathbf{k}$ and $\psi^R_\mathbf{k}$ is the (right) zero mode of $\mathbb{I} - \widehat{C}(0)$ 
evaluated at the transition point. 
This scaling holds when approaching the A3
critical point in any direction other than the one parallel to the line of usual A2 transitions that approach
the A3 point. In this case one finds 
$g_\mathbf{k} \propto \sqrt{\epsilon} \psi^R_\mathbf{k}$. Away from criticality, 
the smallest eigenvalue of the matrix $\mathbb{I} - \widehat{C}(0)$ is not exactly zero: the deviations are of order, respectively, $\epsilon_\perp^{2/3}$ and $\epsilon_\parallel$ close to an A3 point, depending on the direction of approach to criticality. 
Coming back to our original problem, we remark that the eigenvalues of $C_{\mathbf{q};\mathbf{k}}(\mathbf{q}_0)$ can be computed using perturbation theory. Due to rotational invariance, one finds that all eigenvalues of $C_{\mathbf{q};\mathbf{k}}(0)$ are shifted by an amount $\propto q_0^2$. The smallest
eigenvalue of $\mathbb{I} - \widehat C(\mathbf{q}_0)$ is therefore equal to $\alpha |\epsilon_\perp|^{2/3} + \Gamma q_0^2$, where $\alpha$ and $\Gamma$ are
numbers, and $\epsilon_\perp \neq 0$. The solution of Eq. (\ref{chi-eq}) will be dominated by this very small eigenvalue, and thus reads:
\be
\chi_{\mathbf{q}_0}(\mathbf{k},t \to \infty) \approx \frac{\langle \psi^L | S \rangle \psi^R_\mathbf{k}}{\alpha |\epsilon_\perp|^{2/3} + \Gamma q_0^2},
\ee
where $\psi^{L,R}$ are the left and right largest eigenvectors of $C_{\mathbf{q};\mathbf{k}}$ at criticality. From this expression, one
directly demonstrates the existence of a diverging susceptibility and a diverging length scale within MCT, 
which is intimately due to the vanishing of 
the curvature of the FP potential. Close to an A3 point, this length scale diverges as $|\epsilon_\perp|^{-1/3}$, as announced above. The time dependent analysis is more 
cumbersome and will be presented in detail elsewhere \cite{us-long}. The final result is: 
\bea
\chi_{\mathbf{q_0}}(\mathbf{k},t)&\simeq &\frac{V_{\perp}(k)\xi_\perp^{2}}{1 +\Gamma (q_0\xi_\perp)^2} 
\mathcal{G}_\parallel\left(\frac{\ln t}{\sqrt{\xi_{\perp}}},q_0 \xi_\perp\right), \quad \epsilon_\perp \neq 0, \label{xi_scaling_q0}\\ \nonumber
\chi_{\mathbf{q_0}}(\mathbf{k},t)&\simeq& \frac{V_{\parallel}(k)\xi_\parallel^2}{1 +\Gamma (q_0 \xi_\parallel)^2} 
\mathcal{G}_\perp\left(\frac{\ln t}{\sqrt{\xi_{\parallel}}},q_0 \xi_\parallel\right), \quad \epsilon_\perp = 0, \label{eta_scaling}
\eea
where $V_{\parallel,\perp}(k)$ are certain functions, and $\xi_\perp=|\epsilon_\perp|^{-1/3}$, $\xi_\parallel=|\epsilon_\parallel|^{-1/2}$, 
as indeed anticipated by the simple arguments above. 

The scaling functions $\mathcal{G}_{\parallel,\perp}(u,v)$ are such that in the limit $\epsilon_{\parallel,\perp} \to 0$ and $q_0 = 0$ (corresponding
to a uniform perturbation), the dynamical susceptibility is well defined. This imposes $\mathcal{G}_{\parallel,\perp}(u,v=0) \sim u^4$, and therefore, at
criticality, $\chi_{\mathbf{0}}^c(\mathbf{k},t) \sim \ln^4 t$, a result that can be obtained directly by taking the
derivative of the scaling form of $F(\mathbf{k},t)$ close to an A3
critical point, see \cite{Gotze}, and that is compatible with the simulation results of attractive colloids reported in \cite{PCDR}. Away from criticality, the 
$\ln^4 t$ behaviour only persists up to a time $\tau_\xi$ such that $\ln \tau_\xi \sim \sqrt{\xi_\parallel}$ (or $\ln \tau_\xi \sim \sqrt{\xi_\perp}$ along the
special line $\epsilon_\parallel=0$.). Note that the exponential dependence of the relaxation time as a function of the distance from the critical point
was already known from the critical behaviour of the dynamical structure factor, see \cite{Gotze}. Another interesting limit is when the system is critical 
$\epsilon_{\parallel,\perp} = 0$ and perturbed at a non-zero spatial frequency, $q_0 \neq 0$. In order to retain a non-trivial dynamics, one must now have
$\mathcal{G}_{\parallel,\perp}(u \to 0,v \to \infty) = g(u^2 v)$, where $g(x)$ is a certain function behaving as $x^2$ for small $x$, and saturating to
a constant for $x \to \infty$. This leads to: $\chi_{\mathbf{q_0}}^c(\mathbf{k},t) = q_0^{-2} g(q_0 \ln^2 t)$, which shows that at criticality, 
the dynamical length grows without bound, as $\xi_c(t) \sim \ln^2 t$ \cite{F2}. This should be compared with the corresponding result for the A2 point, 
where $\xi_c \sim t^{a/2}$, where $a$ is the MCT exponent for the $\beta$-regime. As the A3 point is approached, the value of $a$ tends to zero and the
power law crosses over to a logarithmic behaviour. Note that there is no $\alpha$-regime at the A3 point, contrarily to the usual phenomenology of the A2 transition. In the latter case, the $q_0$ dependence of $\chi_{\mathbf{q_0}}$ for $q_0 \gg \xi^{-1}$ crosses over from 
$q_0^{-2}$ in the short time, $\beta$-regime, to $q_0^{-4}$ in the long time, $\alpha$-regime. For the A3 transition, on the other hand, only the $q_0^{-2}$ behaviour survives. We have checked all these results numerically by solving exactly the dynamical equation obeyed by $\chi_{\mathbf{q_0}}(t)$ in the so-called schematic limit where all $\mathbf{k}$ dependence is discarded. We have chosen the $F_{13}$ model for which the memory kernel is $\mathcal{F}_{\mathbf{\varepsilon}}[f]=\epsilon_1 f +
\epsilon_3 f^3$. The salient features of the above scaling predictions for $\chi_{\mathbf{q_0}}(t)$ are confirmed in Fig. 2. From a purely phenomenological point of view, the most important points are: (1) the growth 
of $\chi_{\mathbf{q_0}}$ is a logarithmic function of the relaxation time thus implying that dynamical heterogeneities increase much slower close to an A3 critical point than close to an A2 one; this might explain the numerical data of \cite{Jack}; (2) the shape of $\chi_{\mathbf{q_0}}$ is markedly different from the one found at an ordinary MCT transition. In particular, both the maximum and the long-time limit of $\chi_{\mathbf{q_0}}$ diverge as the transition is approached, at variance with what happens close to an A2 point, where the long-time limit of $\chi_{\mathbf{q_0}}$ remains bounded.

All the above results should be only be valid in high enough dimensions. In order to asses the effect of critical fluctuations on mean-field results one has to focus on the 4-point density correlations often called $G_4(\mathbf{r})$, which is related (in Fourier space) to the {\it square} of the dynamical susceptibility 
$\chi_{\mathbf{q}_0}$ defined above (see \cite{JChemPhys} for a full justification of this relation). Since $\chi_{\mathbf{q}_0}$ behaves as $q_0^{-2}$
at criticality, $G_4(r)$ is found to decay as $1/r^{d-4}$ up to distances of order $\xi$. This allows one to estimate the intensive fluctuations of the order parameter 
$f^*$ in a region of size $\xi^d$, which is found to be $\sqrt{\langle \delta f^2 \rangle} \sim \xi^{-(4-d)/2}=
\epsilon^{\nu (4-d)/2}$, which must be compared to $f^* - f_c \sim 
\epsilon^\beta$. Since $\beta=\nu=1/3$ for A3 transitions (or $\beta=\nu=1/2$ along the special direction) we conclude that the upper critical dimension $d_c$ below which critical fluctuations change the
nature of the transition is $d_c=6$, instead of $d_c=8$ as found for usual MCT transitions \cite{JChemPhys,FPFT}. Therefore, in physical dimensions $d=3$, one expects 
that these critical fluctuations will considerably affect the above predictions, at least close enough to the critical point. 
The relative influence of these fluctuations, and the quantitive size of the Ginzburg region where 
critical fluctuations are strong, are expected to depend on the model. 
But one consequence of these fluctuations that is physically relevant 
is the violation of the Stokes-Einstein relation, relating the relaxation time of the system to the diffusion constant of probe particles. 
Simulations of thermal and athermal \cite{Eaves,CPZ} systems appear to conform to the prediction $d_c=8$ 
for usual MCT transitions, yielding, for example, a Stokes-Einstein violation exponent that vanishes linearly as $8-d$ \cite{BBSE}.  
Our above analysis demonstrates that near an A3 singularity the upper critical dimension of fluctuations is shifted down to a value $d_{c}=6$.  
This result is indeed qualitatively consistent with the fact that some aspects of dynamical heterogeneity (such as bimodality of
particle displacement distributions) are suppressed as one tunes, via the introduction of short-ranged attractions,
a supercooled hard-sphere suspension to a regime dominated by higher-order
singularities \cite{Rabani}. More numerical work on this and other aspects of our theory would be welcome, using simulations of attractive hard-spheres in higher
dimensions, along the lines of \cite{CPZ}. 

Finally, the alert reader will have recognized, both from the evolution of the FP potential shown in Fig. 1 and the value of the exponent $\beta=1/3$ 
in generic directions and $\beta=1/2$ in a special direction, that the A3 critical point is akin to an Ising transition, where $\epsilon_\parallel$ is 
a magnetic-field-like perturbation and $\epsilon_\perp$ is a temperature-like perturbation. The behaviour of the correlation function 
$G_4(r) \sim r^{4-d}$ and the corresponding value $d_c=6$ point towards the universality class of the RFIM, in line with previous static treatments \cite{CBRPGT,FPRFIM}. Indeed, the A2 line is analogous to the spinodal line of the RFIM \cite{FPFT}, which terminates
at the A3 RFIM critical point.
Remarkably, the logarithmic behaviour of the correlation function and of 
the dynamical correlation length that we found is usually a manifestation of {\it activated events}, which are indeed expected for the RFIM at criticality, but were thought to be impossible to grasp 
within an MCT formalism \footnote{It is intriguing that MCT predicts near the A3 point a logarithmic behaviour 
for both the correlation function and for the dynamical correlation length, $\xi \sim \ln^2 t$.  
This corresponds to the so-called Sinai diffusion law, which governs the dynamics of
domain walls in the RFIM along the branches of a tree. As noted above, the logarithmic behaviour turns into a power law behaviour $\xi \sim t^{a/2}$ in the presence 
of a field, with a continuously varying exponent, exactly as for the Sinai model. Whether this analogy is accidental or reveals a deep and interesting 
physical feature of the MCT formalism remains to be seen.}. On the other hand, as mentioned above, 
cooperative activated processes not described by MCT are either suppressed or not relevant at an A3 critical point, thus opening the possibility that above $d>6$ MCT indeed captures the correct behaviour. Curiously, however, the dynamical behaviour of the RFIM in high dimensions has not been worked out yet. Usual scaling arguments leading to $\log \tau \propto \xi^\theta$ (where $\theta$ is the stiffness exponent) are expected to break down above $d=d_{DR}\simeq 5.1$ \cite{TT}. Understanding to what extent dynamic scaling holds
above $d_{DR}$ and comparing the actual RFIM dynamical behaviour above six dimension to the prediction of MCT near an A3 point is certainly a topic worth a further studies.


\begin{acknowledgments}
GB and SN acknowledge support from the ERC grant NPRGGLASS. DRR acknowledges NSF CHE-1213247. 
KM acknowledges KAKENHI No. 24340098,25103005, and the JSPS Core-to-Core Program.
We wish to thank C. Cammarota, V. Krakoviack, M. Sperl and G. Tarjus for helpful discussions. 
\end{acknowledgments}


\begin{thebibliography}{99}



\bibitem{sciortino}
F. Sciortino and P. Tartaglia, Adv. in Phys. {\bf 54} 471 (2005).

\bibitem{krakoviack}
V. Krakoviack, Phys. Rev. Lett. {\bf 94}, 065703 (2005); Phys. Rev. E {\bf 75}, 031503 (2007); Phys. Rev. E {\bf 84}, 050501(R) (2011). 

\bibitem{CBRPGT}
C. Cammarota, G. Biroli,  PNAS {\bf 109}, 8850 (2012); J. Chem. Phys. {\bf 138}, 12A547 (2013).

\bibitem{CC}
C. Cammarota, Europhysics Letters {\bf 101} 56001 (2013).

\bibitem{FPRFIM}
S. Franz, G. Parisi, JSTAT (2013) P11012. 

\bibitem{zaccarelli}
K. A. Dawson, G. Foffi, M. Fuchs, W. G\"otze, F. Sciortino, M. Sperl,
P. Tartaglia, Th. Voigtmann and E. Zaccarelli, Phys. Rev. E {\bf 63}, 11401 (2001).


\bibitem{berthierkob}
W. Kob, L. Berthier, Phys. Rev. Lett. {\bf 110}, 245702 (2013).

\bibitem{IMCT}
G Biroli, J.-P. Bouchaud, K. Miyazaki, D.R. Reichman
Phys. Rev. Lett. {\bf 97}, (2006) 195701.

\bibitem{FP}
S. Franz and G. Parisi, J. Phys. (Paris) I {\bf 5}, 1401 (1995).

\bibitem{F1}
The connection between 
the derivative of the FP potential and the large time value of the MCT memory kernel is only qualitative. 
In fact $\frac{f}{1-f}-\mathcal{F}_{\mathbf{\varepsilon}}[f]$  and $V'(f;\mathbf{\varepsilon})$ have the same shape, which is what we need for our discussion, but they are not equal.

\bibitem{Gotze}
W. G\"otze, {\it Complex dynamics of glass-forming liquids}, Oxford Science Publications

\bibitem{FZ}
M. Sellitto, F. Zamponi -Europhys. Lett. {\bf 103}, 46005 (2013). 

\bibitem{FPchi}
S. Franz, G. Parisi, J. Phys. Cond. Matt. {\bf 12} (2000) 6335.

\bibitem{BBchi}
G. Biroli and J.-P. Bouchaud, Europhysiscs Letters {\bf 67}, 21 (2004).

\bibitem{MF}
S. Franz, A. Montanari, Journal of Physics A, {\bf 40} F251 (2007).

\bibitem{kim}
K. Kim, S. Saito, K. Miyazaki, G. Biroli, D. R. Reichman,  J. Phys. Chem. B {\bf 117}, 13259 (2013). 

\bibitem{andreanov}
A. Andreanov, G. Biroli and J.-P. Bouchaud, Europhysics Lett. {\bf 88} 16001 (2009). 

\bibitem{urbani}
S. Franz, H. Jacquin, G. Parisi, P. Urbani, F. Zamponi, PNAS {\bf 109}, 18725 (2012).

\bibitem{PCDR}  P. Charbonneau, D. R. Reichman, Phys. Rev. Lett. {\bf 99}, 135701 (2007) 

\bibitem{F2}
Note that $\xi_c(\tau_\xi) \equiv \xi_{\parallel, \perp}$, in full agreement 
with the above relation between $\tau_\xi$ and $\xi$.

\bibitem{Jack} R. L. Jack, C. J. Fullerton, Phys. Rev. E {\bf 88}, 042304 (2013) 

\bibitem{JChemPhys}
L. Berthier, G. Biroli, J.-P. Bouchaud, W. Kob, K. Miyazaki, D. Reichman, 
J. Chem. Phys. {\bf 126}, 184503 and  184504 (2007).

\bibitem{us-long}
S. Nandi, G. Biroli, J.-P. Bouchaud, K. Miyazaki, D.R. Reichman, in preparation.

\bibitem{FPFT}
S. Franz, G. Parisi, F. Ricci-Tersenghi, T. Rizzo, Eur. Phys. J. E. (2011) {\bf 34} 102.

\bibitem{Eaves}
J. D. Eaves and D. R. Reichman, PNAS {\bf 106}, 15171 (2009).

\bibitem{BBSE}
G. Biroli and J.-P. Bouchaud, J. Phys.: Condens. Matter {\bf 19} 205101 (2007).

\bibitem{Rabani}
D. R. Reichman, E. Rabani and P. L. Geissler, Journal of Physical Chemistry B {\bf 109}, 14654 (2005).

\bibitem{CPZ} 
B. Charbonneau, P. Charbonneau, Y. Jin, G. Parisi, F. Zamponi
 J. Chem. Phys. {\bf 139}, 164502 (2013).
 
 \bibitem{TT}
 M. Tissier, G. Tarjus, Phys. Rev. B {\bf 85}, 104203 (2012).
 

\end{thebibliography}
\end{document}